\documentclass[iop, twocolappendix, appendixfloats, numberedappendix, apj]{hackemulateapj}
\usepackage{graphicx}

\usepackage{latexsym,amssymb}

\usepackage{amsmath,morefloats}
\usepackage[backref,breaklinks,colorlinks,citecolor=blue]{hyperref}
\usepackage{natbib,graphicx,amsmath,subfigure,color,xcolor}
\usepackage{verbatim}
\usepackage{threeparttable}

\newcommand{\galsim}{\texttt{GALSIM}}
\newcommand{\ngmix}{\texttt{ngmix}}

\newcommand{\snr}{$S/N$}

\newcommand{\mcal}{\textsc{metacalibration}}
\newcommand{\mdet}{\textsc{metadetection}}

\newcommand{\Mcal}{\textsc{Metacalibration}}
\newcommand{\Mdet}{\textsc{Metadetection}}

\newcommand{\sx}{\textsc{Source Extractor}}

\newcommand{\bfd}{\textsc{BFD}}

\newcommand{\vonkarman}{{von K\'arm\'an}~}

\shorttitle{\Mdet}
\shortauthors{Sheldon, Becker, MacCrann, Jarvis}

\begin{document}
% mnrad
% \date{Draft \today}

% mnras
%\maketitle

% apj
%\title{\Mdet: Mitigating Shear-dependent Object Detection Biases with \Mcal}
\title{Mitigating Shear-dependent Object Detection Biases with Metacalibration}

\author{Erin S. Sheldon}
\affil{Brookhaven National Laboratory, Bldg 510, Upton, New York 11973, USA}
\author{Matthew R. Becker}
\affil{High Energy Physics Division, Argonne National Laboratory, Lemont, IL 60439, USA}
\author{Niall MacCrann}
\affil{Center for Cosmology and Astro-Particle Physics, The Ohio State University, Columbus, OH 43210, USA}
\affil{Department of Physics, The Ohio State University, Columbus, OH 43210, USA}
\author{Michael Jarvis}
\affil{Department of Physics and Astronomy, University of Pennsylvania, Philadelphia, PA 19104, USA}

\begin{abstract}

    \Mcal\ is a new technique for measuring weak gravitational lensing shear that
    is unbiased for isolated galaxy images.  In this work we test \mcal\ with
    overlapping, or ``blended'' galaxy images.  Using standard \mcal, we find a few
    percent shear measurement bias for galaxy densities relevant for current
    surveys, and that this bias increases with increasing galaxy number density.
    We show that this bias is not due to blending itself, but rather to
    shear-dependent object detection. If object detection is shear independent, no
    deblending of images is needed, in principle.  We demonstrate that detection
    biases are accurately removed when including object detection in the \mcal\
    process, a technique we call \mdet.  This process involves applying an
    artificial shear to images of small regions of sky and performing detection on
    the sheared images, as well as measurements that are used to calculate a shear
    response.   We demonstrate that the method can accurately recover weak shear
    signals even in highly blended scenes.  In the \mcal\ process, the space
    between objects is sheared coherently, which does not perfectly match the real
    universe in which some, but not all, galaxy images are sheared coherently.  We
    find that even for the worst case scenario, in which the space between objects
    is completely unsheared, the resulting shear bias is at most a few tenths of a
    percent for future surveys.  We discuss additional technical challenges that
    must be met in order to implement \mdet\ for real surveys.       

\end{abstract}

\section{Introduction} \label{sec:intro}

Recently developed methods to estimate weak gravitational lensing shear can in
principle provide calibration at the 0.1\% level or better, sufficient for the
requirements of future weak lensing surveys \citep[e.g.,][]{huterer2006}.  At
the time of writing, two methods have demonstrated sufficient accuracy without
reliance on calibration from simulations, including rigorous mathematical
formalisms to deal with selection effects:  the \bfd\ method
\citep{BernBFD2016} and the \mcal\ method \citep{HuffMcal2017,SheldonMcal2017}.
The FPFS method of \cite{FPFS2018} is also able to achieve the required
accuracy using an iterative method to deal with selection effects.  However,
these methods do not deal explicitly with an important aspect of the real
universe: the images of objects overlap on the sky and thus the light from
separate objects is ``blended'' \citep[for discussion of blending
effects see, e.g.,][]{DawsonBlending2016}.

\mcal\ can, in principle, be used calibrate any shear-dependent measurement
biases, even those associated with blending.  However, we will show below that
there is a particular calibration bias associated with the process of detecting
objects in the presence of blending, and this is not addressed in naive
implementations of \mcal. This bias, if left unaddressed, will greatly exceed
the strict requirements for future lensing surveys \citep{huterer2006}, in
which blending will be prevalent \citep{DawsonBlending2016}.

At this stage it is worthwhile to define exactly what we mean by object
detection.  For isolated objects, object detection is closely related to the
identification of regions of an image with pixel values above some threshold.
But when objects overlap on the sky, whether due to physical association or
chance projection, it may be useful to determine how many objects there are in
each detected region. This may be important, for example, when calculating a
redshift distribution for a set of detections.  We associate this process of
identifying the individual objects in a super-threshold region with detection.
We reserve the term ``deblending'' to specifically mean the process of
assigning a fraction of the light in each pixel to each detected object, which
may or may not be a feature of the object detection.  In this work, for the
sake of brevity, we will use the terms ``detection'' and ``object detection''
interchangeably.

\begin{figure*}
    \begin{center}
        \includegraphics[width=\textwidth]{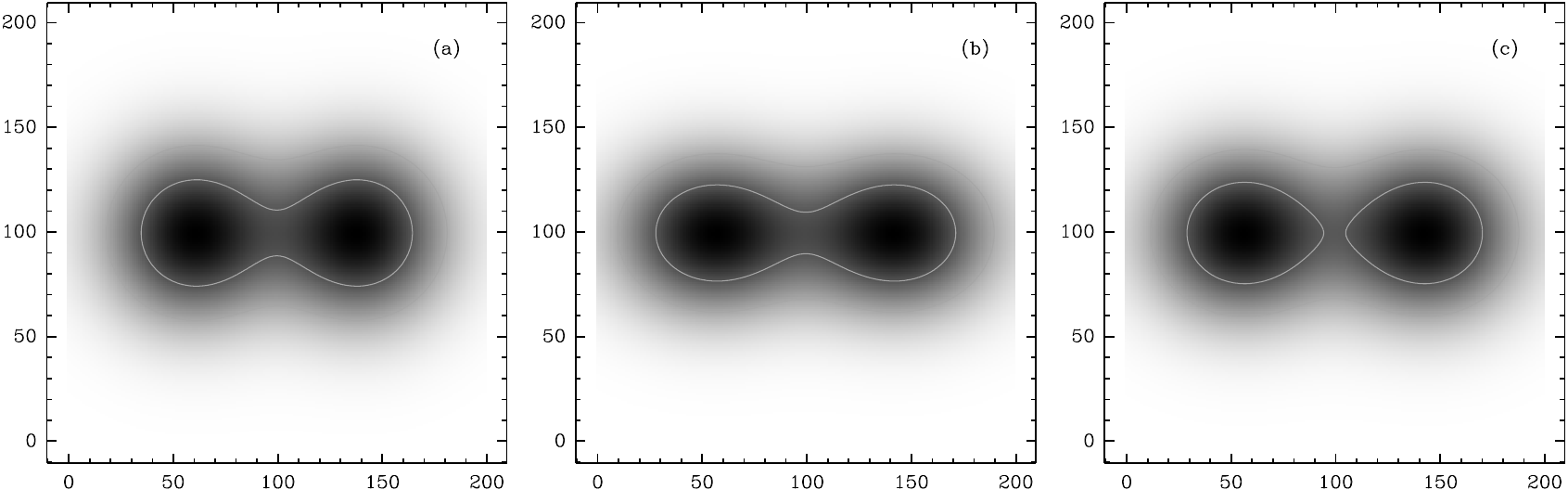}

        \caption{ Toy example of shear-dependent object detection in the presence of
        a PSF.  In panel (a) two objects are present, convolved by a PSF with no
        shear.  Contours represent constant brightness.  In panel (b) the objects
        are sheared by $\gamma = (0.0, 0.1)$ {\em after} the PSF convolution.  The
        contour levels are the same as panel (a).  In this case the inner contours
        for the two objects overlap before and after application of the shear. This
        is a general property of the shear transformation in the weak regime:
        surface brightness is preserved under shear, and because the mapping is
        one-to-one, the topology is also preserved. In panel (c) the shear is
        applied {\em before} the PSF convolution, which mimics real sky images. In
        this case the inner contours do not overlap after shearing, and two objects
        may be detected rather than one.  For case (c) an object detection
        algorithm that identified connected regions above a threshold as a single
        object would manifest a shear-dependent object detection bias.
        \label{fig:toy} }
    \end{center}

\end{figure*}

In the weak shear regime, the lensing mapping is one-to-one and preserves
surface brightness \citep{SchneiderBook92}. For such a mapping, object
detection need not be shear dependent. For example, an object detection
algorithm that finds connected regions in an image with pixel values above a
threshold will not in principle be shear dependent.  This is because the
topology of the contours, or the number of closed contours, at a given surface
brightness will not change under a shear, due to the preservation of surface
brightness and the one-to-one nature of the mapping.  However, in real
observations the image resolution is degraded by a point-spread-function (PSF)
due to the atmosphere, telescope optics and detector. In this case the overlap
of objects, and also the topology of contours of a given surface brightness,
does depend on shear because the PSF convolution occurs after the shear
mapping.  This effect is demonstrated in Figure~\ref{fig:toy}.  Thus the simple
threshold object detection method described above will manifest a
shear-dependent object detection bias.  Note this effect is present even in the
absence of pixel noise, so blended objects well above the detection threshold
will also manifest a detection bias.

% As we will demonstrate, the shear bias caused by this effect far exceeds the
% requirements for future surveys \citep{huterer2006}.  

Common object detection schemes in use today, such at those in \sx\
\citep{Bertin96} and the pipelines used for the Hyper-Surprime Camera survey
and the Rubin Observatory Legacy Survey of Space and Time (LSST)
\citep{BoschHSC2018,BoschLSST2018} are based on thresholding, similar to the
simple approach described above but differing in complexity and efficiency.
The detection and splitting procedure is demonstrated clearly for \sx\ in
\cite{Bertin96} Figure~2.  As we will show, the object detections
produced by \sx\ do manifest shear dependence.  Even a simple local peak
finder, run on a smoothed image, has similar properties and manifests the bias.
An open question, which we will not address in this work, is whether it is
possible to derive a shear-independent object detection algorithm in the
presence of a PSF and a detector with finite spatial resolution.  Such an
algorithm would in principle eliminate the shear-dependent object detection
biases explored in this work.

Published implementations of \mcal\
\citep[e.g.,][]{HuffMcal2017,SheldonMcal2017}, when used with the common object
detection schemes discussed above, are expected to exhibit shear-dependent
object detection biases. These implementations of \mcal\ work by applying
artificial shears to small ``postage stamp'' images. These postage stamps are
extracted from a larger image at the locations of objects found during an {\em
independent} object detection step, run before the application of \mcal.
Measurements on the sheared images are then used to calculate of a linear
response of an ellipticity measurement to the applied shear. Because the
independent object detection already manifests a shear-dependent object
detection bias, the shear applied when running \mcal\ does not properly reflect
the full response to shear.

% Shear-dependent object detection bias is a type of selection effect.
% Corrections for selection effects in a static catalog were already derived in
% \cite{SheldonMcal2017}. In that formalism one applies selections to quantities
% measured on artificially sheared images, and recalculates the shear estimate to
% derive a selection shear response.  This selection response is then added
% to the shear response.  However, that method assumes that the base
% catalog itself is itself unaffected by shear-dependent selection effects.
% Similarly, that formalism cannot work for objects near the object detection
% threshold, even for isolated objects, because the object detections needed for
% the corrections will, by definition, not be present in the catalog.

Object detection effects can be naturally incorporated into \mcal\ by shearing
larger images, rather than small postage stamps, and re-running the object
detection algorithm on each of the sheared images.
% In this process, ellipticity
% measurement biases, selection effects and object detection are all addressed.
We call this technique \mdet\ to indicate that detection is part of the \mcal\
process.  An rough outline of the method is as follows: \begin{enumerate}

    \item Artificially shear images of small regions of sky.

    \item Perform detection and measurement on the sheared images, as well as
        images without artificial shear.  Due to shear-dependent detection
        effects, the number of detections found the each image will be slightly
        different, resulting in different object catalogs.

    \item Calculate ensemble shear statistics using measurements from the
        unsheared images.

    \item Use measurements from sheared images to calculate a linear response
        of the measurements to a shear.

    \item Correct the ensemble shear statistics using the response.

\end{enumerate}
The ensemble mean ellipticities are taken over different catalogs, whereas in
\cite{SheldonMcal2017} all measurements were performed using a single static
detection catalog.  

As we will demonstrate, \mdet\ can correct for biases associated with
shear-dependent detection, in addition to model bias, pixel noise bias
and ordinary shear-dependent selection effects.

An alternative route being explored by the community is to calibrate shear
measurements using simulations \citep[see, e.g.,][]{KIDS450shear,HSCY1shear}.
The idea is to use an uncalibrated ellipticity measurement, which can manifest
a large shear bias (of order 10\%) and  unknown selection and detection
effects.  The final shear measurements are then calibrated using a simulation
that matches the data as closely as possible.  Ultimately \mdet\ may have some
remaining biases that must be calibrated with simulations.  Our goal in
developing this algorithm is to make these biases as small as possible,
sub-percent as opposed to 10 percent, which we expect will greatly reduce the
sensitivity of the final calibration to small inaccuracies in the simulation.

The paper is laid out as follows. In Section~\ref{sec:mdet} we introduce a
generalized \mcal\ that accounts for shear dependent object detection.  In
Section~\ref{sec:sims}, we describe the simulations we use to test our methods
and the analysis techniques used to infer the shear signal. In
Section~\ref{sec:detbiases}, we study the effects of object detection on shear
measurements with \mcal.  In Section~\ref{sec:mitigate}, we describe in detail 
our implementation of \mdet\ and apply it to recover the shear signal
in our simulations. We also discuss the physical limits of
\mdet.  In Section~\ref{sec:psfvar}, we study the effects of the PSF
variation on \mdet.  Finally, we conclude in Section~\ref{sec:conc}.

\section{\textsc{Metacalibration} and \textsc{Metadetection}}
\label{sec:mdet}

In this section we introduce an extension of \mcal\ that naturally accounts for
shear-dependent object detection.

\mcal\ is a general technique that computes the linear response of measurements
on an image to an applied shear using only the observed image.  For a small
applied shear $\boldsymbol{\gamma}$, we can expand a two-compoment
measurement $\boldsymbol{e}$ as
\begin{eqnarray}
\boldsymbol{e} & \approx & \left.\boldsymbol{e}\right|_{\gamma=0} +
                           \left.\frac{\partial \boldsymbol{e}}{\partial\boldsymbol\gamma}\right|_{\gamma=0} \boldsymbol\gamma +
                           O(\boldsymbol\gamma^2)\nonumber\\
               & \equiv  & \left.\boldsymbol{e}\right|_{\gamma=0} +
                           \boldsymbol{R} \boldsymbol\gamma +
                           O(\boldsymbol\gamma^2)
\end{eqnarray}
where $\boldsymbol{R}$ is the response matrix of the image measurement
at zero applied shear, $R_{ij}=\partial e_i /\partial \gamma_j$, with $i$ and
$j$ taking all combinations of the two shear components.  The measurement
$\boldsymbol{e}$ can be, for example,  an ellipticity measurement, and in what
follows we will use the term ellipticity without loss of generality.

We estimate the response using a numerical, finite-difference derivative
\begin{equation}
R_{ij} \approx \frac{e_i^{+} - e_i^{-}}{\Delta\gamma_j}\ .
\end{equation}
where $R_{ij}$ is the estimated response of the measurement to shear and
$\Delta\gamma_j = 2 \gamma_j$ is the difference between two applied shears,
$\pm \gamma_j$, with $\gamma_j$ a small shear of order 0.01. The quantity
$e_i^{+/-}$ is the $i$-th ellipticity component measured on an image sheared
with $\pm\Delta\gamma_j$.  The creation of artificially sheared images requires
careful handling of the PSF and other observational effects
\citep{SheldonMcal2017}.

Because the estimated responses are noisy for a single object, we average the
response over many images and objects.

Let's examine the case of estimating the mean shear from a 
mean ellipticity.  In \cite{SheldonMcal2017} we wrote the mean ellipticity as
\begin{equation}
    \left< \boldsymbol{e} \right> \approx \left< \frac{\partial \boldsymbol{e} }{\partial \boldsymbol{\gamma} } \biggr\rvert_{\gamma=0} \boldsymbol{\gamma} \right>
    = \left< \boldsymbol{R} \boldsymbol{\gamma} \right>
\end{equation}
The mean shear was then estimated using the mean response:
\begin{eqnarray}
    \left< \boldsymbol{R} \right> &=& \left< \frac{\partial \boldsymbol{e} }{\partial \boldsymbol{\gamma} } \biggr\rvert_{\gamma=0} \right>, \nonumber \\
    \langle R_{ij}\rangle &=& \left< \frac{e_i^{+} - e_i^{-}}{\Delta\gamma_j} \right>, \nonumber \\
    \langle \boldsymbol\gamma \rangle & \approx & \langle \boldsymbol{R}\rangle^{-1}\langle\boldsymbol{e}\rangle.
\end{eqnarray}
Note that the derivative appears {\em inside} the averages, implying
we could take the derivative for each object separately.
This is possible for a single static catalog: postage stamp images are generated
for each object, artificially sheared, and  measurements $\boldsymbol e$
are then produced for the sheared images to calculate the response for
each object.

As discussed in \S \ref{sec:intro}, the detection of objects is
shear-dependent.  In particular, the number of objects detected depends on
shear, and we wish to accurately account for this effect in the response.  To
address this we will repeat the detection on artificially sheared images, which
will result in detection catalogs that differ for each sheared image. There is
no single static catalog with which to generate postage stamps and make
individual response measurements.

We can address this by moving the derivative outside of the average:
\begin{eqnarray} \label{eq:fullR}
    \left< \boldsymbol{R} \right> &=& \frac{\partial \left< \boldsymbol{e} \right> }{\partial \boldsymbol{\gamma} } \biggr\rvert_{\gamma=0},  \nonumber \\
    \langle R_{ij}\rangle &=& \frac{\langle e_i^{+}\rangle - \langle e_i^{-}\rangle}{\Delta\gamma_j} \nonumber \\
\end{eqnarray}
The finite difference is now calculated over averages, based on detections 
derived from the different sheared images.  In general, only a subset of the
detected objects will be common to all detection catalogs.
Note that Equation \ref{eq:fullR} is entirely equivalent to the response
Equation 14 of \cite{SheldonMcal2017} in the case of a static catalog.
We call this modified version of the method \mdet.

In what follows we will use simulations to test \mcal\ with and without
including detection in the process.  We will demonstrate that, for standard
detection algorithms, including detection is required to obtain accurate shear
estimates when galaxies are blended.

\section{Analysis and Simulation Techniques}
\label{sec:sims}

In this section, we describe our object simulation and measurement techniques.
In all cases, we used the \galsim\ \citep{GALSIM2015} software package to
generate images, perform convolutions etc. We used the \texttt{SEP} \citep{sep}
Python wrapper of the \sx\ software package \citep{Bertin96} for source
detection as needed. Finally, we used the \ngmix\ software
package\footnote{\url{https://github.com/esheldon/ngmix/}} for object
measurement and the \mcal\ implementation.

\subsection{Multi-object Fitting Deblending} \label{sec:mof}

Multi-object Fitting (MOF) deblending is a technique employed by the Dark
Energy Survey to account for blending of objects when performing image
measurements \citep{DESY1cat}. It is representative of a set of techniques that
involve fitting models to images for a list of preexisting detections. The
model fit is then used directly to form a flux measurement or indirectly by
using it to approximately remove the light of the neighboring objects in the
image before further processing.

In this work, we used an \ngmix\ based MOF
algorithm\footnote{\url{https://github.com/esheldon/mof/}}. It is an improved
version of the MOF fitter used in \cite{DESY1cat} which is both more stable and
faster. It uses a linear combination of a De Vaucouleurs' \citep{devauc1948}
profile and exponential profile. The profiles are constrained to be cocentric,
coelliptical, and to have a fixed one-to-one size ratio.  The relative
amplitude of the two profiles, the fraction of the total flux in the De
Vaucouleurs' profile, is a free parameter in the model. Finally, to process a
large number of objects, we followed \citet{DESY1cat} and broke them up into
associated groups.  These groups of objects were then simultaneously fit using
a least-squares loss function.  When using MOF with \mcal, we created postage
stamp images centered on each object and subtracted the light of neighbors
using the MOF models.

\subsection{Galaxy Pair Simulations}
\label{sec:sims:pairs}

In order to isolate the effects of detection, we employed a simulation setup
consisting of two galaxies.   By varying the separation between the two
galaxies we could carefully tune the effects of blending.

The simulated galaxies were each a combination of a bulge component, modeled as
a De Vaucouleurs' profile \citep{devauc1948} and disk component modeled as an
exponential. The fraction of light in the bulge was random and ranged uniformly
between 0.0 and 1.0. The disk ellipticity was drawn from the distribution
presented in \cite{ba14}, equation 24, with ellipticity variance set to 0.20,
with a random orientation. The bulge was given the same orientation as the disk
but with ellipticity set to the disk ellipticity times a random number drawn
uniformly between 0.0 and 0.5. The half-light radius of the disk
$r_{50}^{\mathrm{disk}}$ was set to a uniform random draw between 0.4 and 0.6
arcsec. The half-light radius of the bulge was a random draw between $0.4
r_{50}^{\mathrm{disk}}$ and $0.6 r_{50}^{\mathrm{disk}}$. The bulge was shifted
from the center of the disk within a radius 0.05$r_{50}^{\mathrm{disk}}$ and in
a random direction. The light of the disk was divided between a smooth
component and a set of simulated ``knots of star formation'', represented by
point sources placed randomly with the same exponential distribution as the
disk.  Between 1 and 50 knots were placed, such that the fraction light in the
knots ranged between 0.4\% and 20\%. The total flux and noise were set such
that the signal-to-noise ratio ranged uniformly between 25 and 35.  The models
were convolved by a PSF modeled as a Moffat \citep{Moffat1969} profile with
$\beta=2.5$ and full with at half maximum 0.9 arcseconds, and rendered into an
image with pixel scale 0.263 arcseconds.

\begin{figure*}
    \begin{center}
        \includegraphics[width=\textwidth]{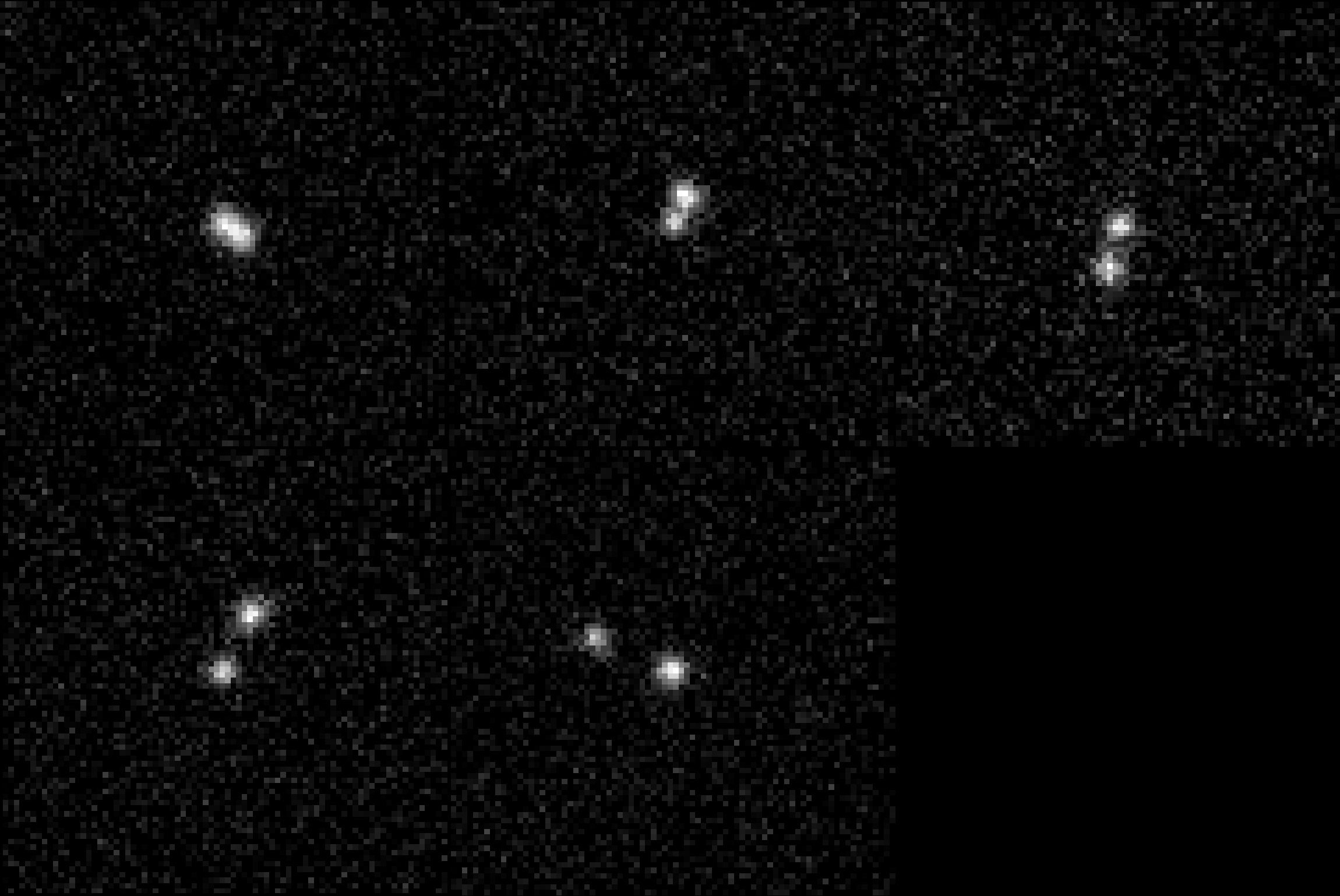}
        \caption{Example images of simulated galaxies used for the pair tests
        presented in section \ref{sec:sims:pairs}.  From left to right in the top row,
        the separations are 1.0, 1.5, 2.0 arcsec. From left to right in the bottom row the
        separations are 3.0 and 4.0 arscec. The pixel scale is 0.263 arcsec.
        \label{fig:pairs}}
    \end{center}
\end{figure*}

We rendered two of these randomly generated galaxies in an image, with
separation ranging from 1.0 and 4.0 arcsec. The pair was situated such that the
line between the pair had a uniform random orientation relative to the
coordinate axes. Each object was given an additional random dither within a
pixel. The galaxies were treated as transparent, such that the value in a pixel
was equal to the total sum from both galaxies plus noise. Example images are
shown in Figure~\ref{fig:pairs}

\subsection{Simulations with Representative Galaxy Density and Noise}
\label{sec:sims:realgals}

In this section we describe simulations with parameters chosen to mimic DES and
LSST images.  We used the publicly available \texttt{WeakLensingDeblending}
software package
\citep{WeakLensingDeblendingSoftware,WeakLensingDeblendingPaper}\footnote{\url{https://github.com/LSSTDESC/WeakLensingDeblending}}
to generate galaxy catalogs and images properties.

We generated images in the r-, i-, and z-bands with an effective depth that is
roughly equivalent to full 5 and 10 year coadd image for the DES and LSST
respectively. For our primary tests, we neglected the effects of PSF variation
and used a constant PSF per-band with the typical (expected) seeing for each
survey ($\sim\!1$ arcsec and $\sim\!0.8$ arcsec respectively). We tested
variable PSFs separately, as discussed in \S \ref{sec:psfvar}.  For the DES
image simulations, we modified the settings slightly such that the effective
exposure time was equivalent to ten 90 second exposures. The depths of the LSST
images were set to match those assumed by the \texttt{WeakLensingDeblending}
package for the 10 year survey, although note the modifications below. In all
simulations a shear of $\gamma_1 = 0.02$ was used.

We made two additional modifications to the simulations produced by this
package.  The galaxy models and PSFs were generated using the
WeakLensingDeblending package, using its internal survey settings and object
catalogs.  But rather than using the package to render into an image, we
rendered them separately so that we could control whether the full scene was
sheared, including the space between objects, or just the objects were sheared
(the only mode supported by WeakLensingDeblending).

\begin{figure*}
    \begin{center}
        \includegraphics[width=0.9\columnwidth]{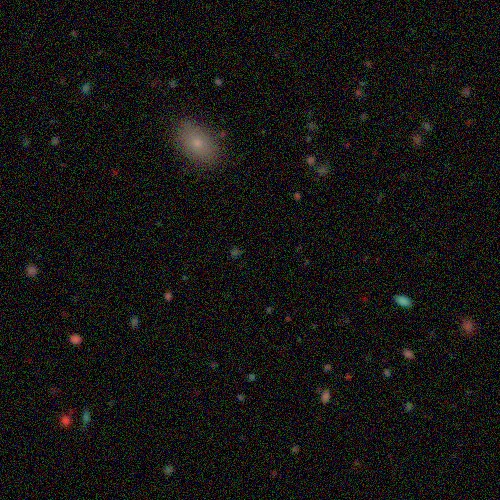}
        \includegraphics[width=0.9\columnwidth]{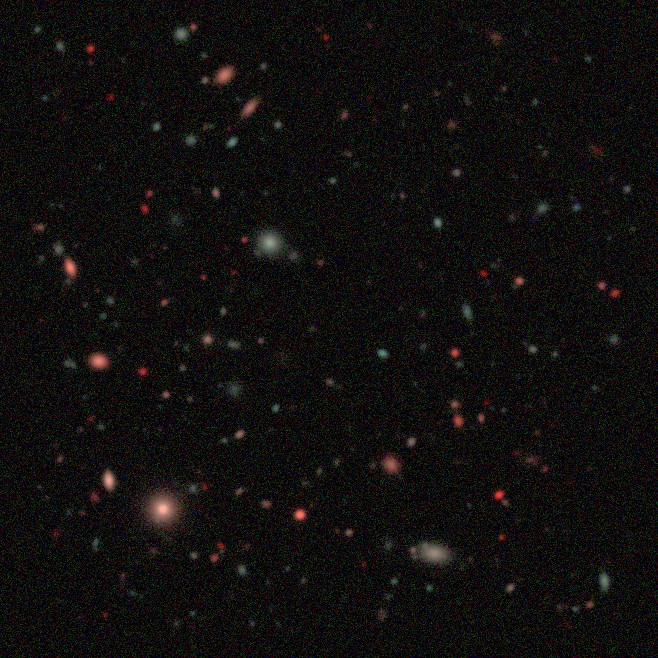}
        \caption{
            Example images from the DES (left) and LSST (right) simulations. Each
        multicolor, $gri$-band image is approximately $\sim\!2.2$ arcmin on a side. The
        DES images have a pixel scale 0.263 arcsec and a PSF FWHM of $\sim\!1$ arcsec.
        The LSST images have a pixel scale of 0.2 arcsec and a PSF FWHM of $\sim\!0.8$
        arcsec.
        \label{fig:simimages}}
    \end{center}
\end{figure*}

We also multiplied the density of input sources by a factor of 0.45 and a
factor of 0.4 for the DES-like and LSST-like simulations, respectively, in
order to produce a realistic number density of detected sources. We also
performed tests with the unmodified input catalog for the LSST-like
simulations, in order to test with an extreme density. The detected source
densities in these three simulations were approximately 35 per square arcmin
for DES year 5, 75 per square arcmin for LSST year 10, and 140 per square
arcmin with the unmodified source catalog for the LSST-like simulations.
Example images for each survey are shown in Figure~\ref{fig:simimages}.

\subsection{Measuring Shear Biases}

We report our results in terms of the standard parameterization of shear
measurement biases \citep[see, e.g.,][]{heymans2006}

\begin{equation} \label{eq:m}
g \equiv c + (1 + m)\gamma
\end{equation}
where $g$ is the recovered shear, $c$ is the additive bias, and $m$ is the
multiplicative bias. Below we report only $m$, but we have found that $c$ is
consistent with zero in all cases.

We used the technique of \citet{pujol2019} to reduce the noise on measurements
of $m$ in our simulations.  This technique works as follows. We generated pairs
of images with identical galaxy and noise properties, but with opposite true
shears applied $\gamma_{\pm}$ (we use a scalar notation because the shear
was only applied in one component).  For each pair of simulations, we calculate the mean
ellipticity and mean response.  We then calculate ensemble means $\langle e_\pm \rangle$
and $\langle R \rangle = ( \langle R_+ \rangle + \langle R_- \rangle)/2$ over all such
pairs of images, and form a difference of the overall recovered shear that
partially cancels noise:
\begin{equation}
    \gamma_{est} = \frac{ \langle e_+ \rangle - \langle e_- \rangle}{2 \langle R \rangle},
\end{equation}
We then calculated $m$ using equation \ref{eq:m}, $\gamma_{est}$ and the true
input shear.  We estimated the errors on the mean $m$ using bootstrap
resampling of the set of image pairs, repeating the computation of $m$ for each
bootstrap sample. We employed a similar procedure for $c$, except that we used
the average of the two estimated shears so that any additive biases common to
both simulations do not cancel. In all  cases we measured $m$ using the
1-component of the shear and $c$ with the 2-component of the shear, though we
have found this choice not to matter in explicit testing.

Note that, in the case that the entire scene is sheared, including the space
between objects, the technique of \citet{pujol2019} does not cancel the noise
as effectively as when only the shapes of objects are sheared.  This is because
the objects move differently under the two different shears and thus sample
slightly different parts of the noise field.

\section{Shear-dependent Detection Biases}\label{sec:detbiases}

In this section, we present results from a set of experiments using image
simulations designed to elucidate the role of detection biases in \mcal\ shear
measurements.  We first examine shear measurement on pairs of galaxies at
various separations.  We then study detection biases in DES- and LSST-like
simulations with realistic galaxy densities and pixel noise. We find in all
cases that object detection imparts a significant shear measurement bias. As we
will show in \S \ref{sec:mdetpairs}, we can correct this bias by including
detection in the \mcal\ process, even if no explicit deblending (division of
light between objects) is performed.

\subsection{Basic Analysis Method}

The full \mdet\ analysis method, which includes detection in the \mcal\
process, was outlined in \S \ref{sec:intro} and the formalism for calibration
was given in \S \ref{sec:mdet}.  In the sections \S
\ref{sec:pairbias} and \S \ref{sec:realbias} that follow, we show shear
recovery bias using standard \mcal, without the full \mdet.   Rather than shearing
a large image and repeating detection, we ran detection once to produce a
static detection catalog.  We then performed basic \mcal\ on postage stamps for
each detection.  During the measurement phase, we performed deblending using
the MOF algorithm described in \S \ref{sec:mof}.  We also ran tests without
deblending, using simple weighted moments, and saw biases at a similar level,
but do not show the results for the sake of brevity.

\subsection{Bias in Simulations of Galaxy Pairs} \label{sec:pairbias}

We tested \mcal\ with using the galaxy pair simulation presented
in \S \ref{sec:sims:pairs}. We used \sx\ for object detection, with settings
matching those used for DES year 5 survey reductions (DES Collaboration, in
prep.)\footnote{We have created a software package to run detection with DES
year 5 settings, using the sep \sx\ wrapper
\url{https://github.com/esheldon/sxdes}}.  We got similar results using a
simple local peak finder for
detection\footnote{\url{https://github.com/esheldon/peaks}}.  We also saw
similar levels of bias with and without performing deblending using MOF.

\begin{figure}
    \includegraphics[width=\columnwidth]{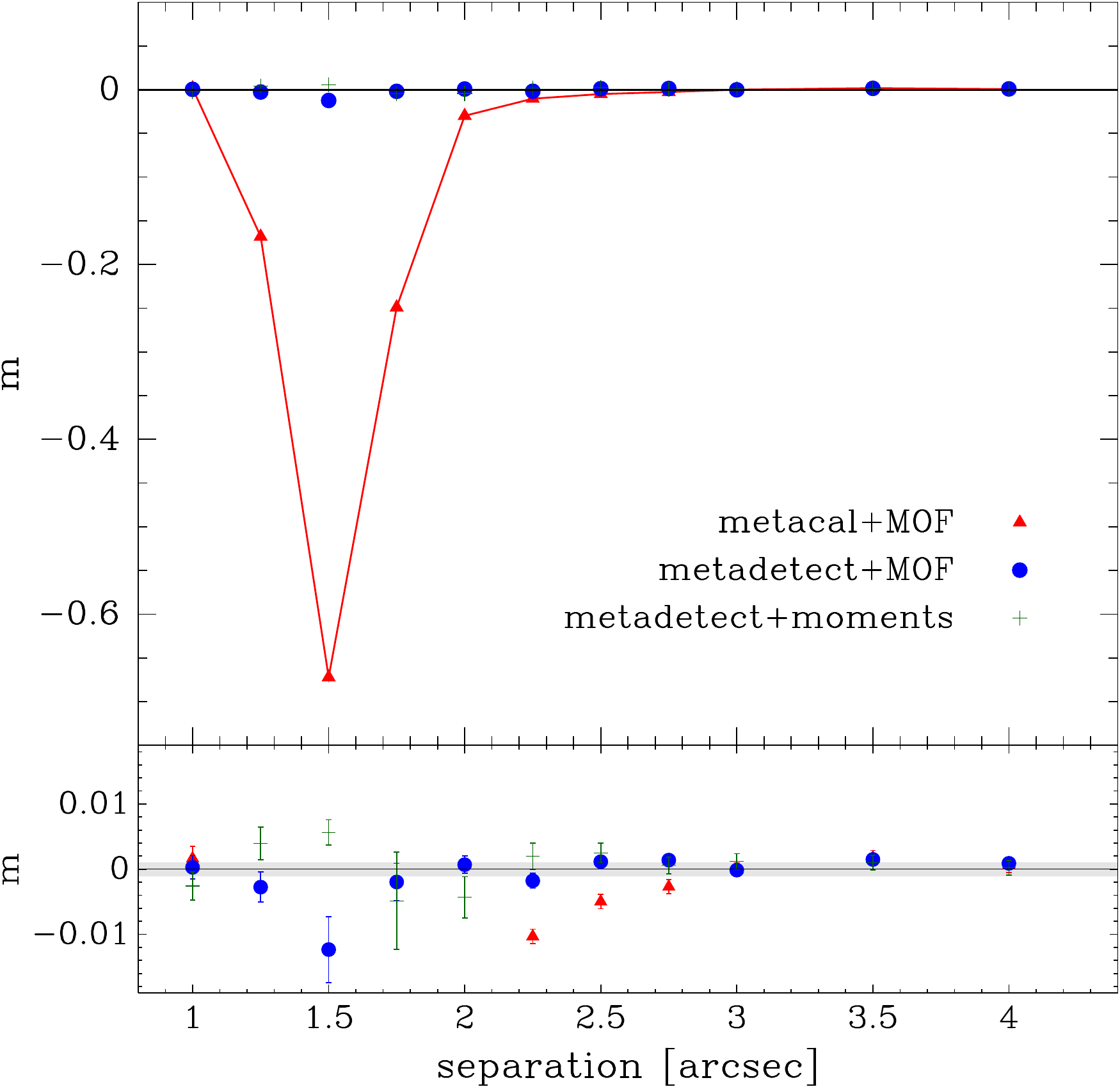}

    \caption{Mean multiplicative shear bias measured for pairs of simulated
    galaxies (see \S \ref {sec:sims:pairs} for details) at various separations.  At
    each separation, a large number of trials was generated with random
    orientations of the pair.  At 4.0 arcsec separation, two objects were
    detected in all cases.  At 1.5 arcseconds two objects were detected in half
    the cases.  At 1.0 arcsec a single object was detected in all cases.  Red
    triangles represent standard \mcal\ with MOF deblending for modeling all
    detected objects.  Blue circles represent \mcal+MOF with detection included
    as part of the process.  Green pluses represent \mcal\ with detection
    included but without deblending, and using simple weighted moments without
    PSF correction as the shear estimator. Very large biases are seen for
    standard \mcal+MOF as detection becomes ambiguous, for example at 1.5
    arcsec separations.  When detection is included in the \mcal\ process the
    biases are greatly reduced.  The bias is reduced even in the case where no
    deblending was performed and no PSF correction or detailed object modeling
    were performed.  This indicates that a majority of the bias is due to
    shear-dependent detection, not light blending or details of the object
    modeling.
    \label{fig:pairbias}}

\end{figure}

The multiplicative bias $m$ is shown in Figure~\ref{fig:pairbias} as a function
of the pair distance. For a large separation of 4 arcsec, two objects were
detected in essentially all cases, but as the separation was decreased the
detection became more ambiguous, with only one object detected in some cases.
At 1.5 arcsec separation the detection was most ambiguous, with two objects
detected in half the cases. As the separation was decreased further, one object
was detected more often than two, and at 1.0 arcseconds only one object was
detected in essentially all cases.  For close separations the blend is
unrecognized but the detection is unambiguous, in the sense that the
detection algorithm consistently finds one object.

In the cases where the detection is unambiguous, at close and far separations,
there is no bias in the recovered shear.  But the bias increases as the
detection becomes more ambiguous. The maximum bias occurs at 1.5 arcsec
separation, where two objects are detected in half the cases, the separation of
maximum ambiguity.

The correspondence between detection ambiguity and shear bias is a hint that
the bias is caused by shear-dependent detection. Next, we test with simulations
of DES- and LSST-like surveys and show explicitly that by making object
detection shear independent, we can eliminate these detection biases.

\subsection{Bias in Simulations with Representative Galaxy Density and Noise}
\label{sec:realbias}

The results for simulations with representative galaxy density and noise are
shown in Table~\ref{tab:shearmeas}.  \mcal\ with MOF deblending and \sx\
detections results in large biases, $\sim-3\%$, for a full five year
DES-like survey. Simulations of LSST-like surveys demonstrate somewhat larger
biases, and a simulation with double the LSST galaxy density shows a very large
bias.  Note, we see similar biases when no deblending is performed but we again
omitted them from the table for the sake of brevity.

The bias numbers presented above are relatively noisy, but it is interesting to
note that the bias for LSST year 10 is not very much larger than DES year 5,
despite the fact that the galaxy density is about twice as high.  This may be
partly due to the better resolution of the LSST images: the area of the LSST
PSF is 60\% smaller than DES PSF.  Thus the images of small galaxies, at fixed
density, will overlap less in LSST images than they do in DES images.

\begin{table*}
  \centering
    \begin{threeparttable}
  \caption{
    Multiplicative biases in weak lensing simulations for various shear
    measurement techniques. In all cases, the simulations use realistic
    galaxy ellipticities, galaxy sizes and noise for the given survey. For measurements using standard \mcal\ with
    MOF deblending, a cut of $T/T_{PSF} > 0.5$ was also applied. Measurements with
    \mdet\ and moments used a size cut of $T/T_{PSF} > 1.2$. In the case of \mdet\ with moments,
    no deblending corrections are applied and the moments are a simple weighted moment
    with no PSF correction.}
  \label{tab:shearmeas}

  %\begin{tabular}{|l|l|l|c|c|}
  \begin{tabular}{lllcc}
    \hline
    \noalign{\vskip 1mm}
    Simulation & Method & Full Scene Sheared? & \snr\ Cut & m \\
    \noalign{\vskip 1mm}
    \hline
    \noalign{\vskip 1mm}

    %\hline
    \multicolumn{5}{c}{metacal+MOF - full scene sheared}\\
    \noalign{\vskip 1mm}
    \hline
    \noalign{\vskip 1mm}
    DESY5   & metacal+MOF & yes & \snr$ > 10$ & $-0.036 \pm 0.005$  \\
    DESY5   & metacal+MOF & yes & \snr$ > 15$ & $-0.023 \pm 0.004$  \\
    DESY5   & metacal+MOF & yes & \snr$ > 20$ & $-0.015 \pm 0.004$  \\
    \noalign{\vskip 1mm}
    \hline
    \noalign{\vskip 1mm}
    LSSTY10  & metacal+MOF & yes & \snr$ > 10$ & $-0.035 \pm 0.002$  \\
    LSSTY10  & metacal+MOF & yes & \snr$ > 15$ & $-0.031 \pm 0.002$  \\
    LSSTY10  & metacal+MOF & yes & \snr$ > 20$ & $-0.026 \pm 0.002$  \\
    \noalign{\vskip 1mm}
    \hline
    \noalign{\vskip 1mm}
    LSSTY10 2$\times$ dens. & metacal+MOF & yes & \snr$ > 10$ & $-0.082 \pm 0.005$  \\
    LSSTY10 2$\times$ dens. & metacal+MOF & yes & \snr$ > 15$ & $-0.067 \pm 0.005$  \\
    LSSTY10 2$\times$ dens. & metacal+MOF & yes & \snr$ > 20$ & $-0.062 \pm 0.004$  \\
    \noalign{\vskip 1mm}
    \hline
    \noalign{\vskip 1mm}

    %\hline
    \multicolumn{5}{c}{metadetect+moments - full scene sheared}\\
    \noalign{\vskip 1mm}
    \hline
    \noalign{\vskip 1mm}
    DESY5   & metadetect+moments & yes & \snr$ > 10$ & $+0.00025 \pm 0.00088$  \\
    DESY5   & metadetect+moments & yes & \snr$ > 15$ & $-0.00085 \pm 0.00070$  \\
    DESY5   & metadetect+moments & yes & \snr$ > 20$ & $+0.00024 \pm 0.00061$  \\
    \noalign{\vskip 1mm}
    \hline
    \noalign{\vskip 1mm}
    LSSTY10  & metadetect+moments & yes & \snr$ > 10$ & $+0.00084 \pm 0.00061$  \\
    LSSTY10  & metadetect+moments & yes & \snr$ > 15$ & $-0.00001 \pm 0.00047$  \\
    LSSTY10  & metadetect+moments & yes & \snr$ > 20$ & $+0.00042 \pm 0.00039$  \\
    \noalign{\vskip 1mm}
    \hline
    \noalign{\vskip 1mm}
    LSSTY10 2$\times$ dens. & metadetect+moments & yes & \snr$ > 10$ & $+0.00047 \pm 0.00043$  \\
    LSSTY10 2$\times$ dens. & metadetect+moments & yes & \snr$ > 15$ & $-0.00002 \pm 0.00034$  \\
    LSSTY10 2$\times$ dens. & metadetect+moments & yes & \snr$ > 20$ & $+0.00019 \pm 0.00028$  \\
    \noalign{\vskip 1mm}
    \hline
    \noalign{\vskip 1mm}

    %\hline
    \multicolumn{5}{c}{metadetect+moments - individual objects sheared}\\
    \noalign{\vskip 1mm}
    \hline
    \noalign{\vskip 1mm}
    DESY5                 & metadetect+moments & no & \snr$ > 10$ & $-0.0042 \pm 0.0009$  \\
    DESY5                 & metadetect+moments & no & \snr$ > 15$ & $-0.0052 \pm 0.0006$  \\
    DESY5                 & metadetect+moments & no & \snr$ > 20$ & $-0.0059 \pm 0.0006$  \\
    \noalign{\vskip 1mm}
    \hline
    \noalign{\vskip 1mm}
    DESY5 2$\times$ dens. & metadetect+moments & no & \snr$ > 10$ & $-0.0029 \pm 0.0006$  \\
    DESY5 2$\times$ dens. & metadetect+moments & no & \snr$ > 15$ & $-0.0017 \pm 0.0005$  \\
    DESY5 2$\times$ dens. & metadetect+moments & no & \snr$ > 20$ & $-0.0023 \pm 0.0005$  \\
    \noalign{\vskip 1mm}
    \hline
    \noalign{\vskip 1mm}
    LSSTY10                 & metadetect+moments & no & \snr$ > 10$ & $-0.0015 \pm 0.0007$  \\
    LSSTY10                 & metadetect+moments & no & \snr$ > 15$ & $-0.0013 \pm 0.0006$  \\
    LSSTY10                 & metadetect+moments & no & \snr$ > 20$ & $+0.0001 \pm 0.0005$  \\
    \noalign{\vskip 1mm}
    \hline
    \noalign{\vskip 1mm}
    LSSTY10 2$\times$ dens. & metadetect+moments & no & \snr$ > 10$ & $-0.0047 \pm 0.0006$  \\
    LSSTY10 2$\times$ dens. & metadetect+moments & no & \snr$ > 15$ & $-0.0035 \pm 0.0004$  \\
    LSSTY10 2$\times$ dens. & metadetect+moments & no & \snr$ > 20$ & $-0.0029 \pm 0.0004$  \\
    \noalign{\vskip 1mm}
    \hline
  \end{tabular}

    \end{threeparttable}
\end{table*}

In order to unpack the source of the bias in this case, we compared two
different \mcal\ shear measurements. The first was performed on a catalog of
the true source positions using a fixed 1.2 arcsecond Gaussian weighted moment
ellipticity measurement. We did not run MOF deblending on the true detection
catalog.  The true catalog is very dense, which means the blended groups would
be very numerous and would have many members.  Because there is a large
computing overhead associated with the deblending, the running time would have
been prohibitively slow.

The second measurement employed \mcal\ with the same weighted moment
ellipticity measurement, but using \sx\ detections rather than the true object
positions.

We found that, while the \mcal\ shear measurement using true detections is
unbiased ($-0.0011\pm0.0012$), the measurement using \sx\ detections exhibits a
bias of $-0.058\pm0.001$. Note that this ellipticity measurement makes no
corrections for object blending, but in the case where we use the true object
positions, it is still unbiased. Thus we have demonstrated that, given a set of
true source locations, \mcal\ is unbiased when objects are blended.

This set of tests also demonstrates explicitly that source detection can cause
significant shear measurement biases even for techniques which are robust to
blending. The source detection biases probably originate from multiple causes,
but one of those causes is certainly the merging or splitting of object detections
in a way that is shear dependent, as illustrated with the toy example in
Figure~\ref{fig:toy} and the galaxy pair tests above.

\section{Mitigating Shear-dependent Detection Biases} \label{sec:mitigate}

In the previous section, we demonstrated that source detection is a significant
source of bias in shear measurements with \mcal. Here we show we can mitigate
this bias by including source detection in the \mcal\ process.  For these
tests, we use the full \mdet\ algorithm outlined in \S \ref{sec:intro}, which
involves first producing large sheared versions of the image, followed
by running detection separately on each sheared image.

We performed object detection on each image using \sx.  We then made
measurements in postage stamps around each detection in each image using a
non-PSF corrected, Gaussian-weighted moment.  We used these five catalogs 
to calculate a single estimate for the shear using the mean shear
response for the image as shown in \ref{eq:fullR} from \S
\ref{sec:mdet}, which we repeat here for clarity:
\begin{eqnarray}
\langle \boldsymbol\gamma \rangle &\approx& \langle \boldsymbol{R}\rangle^{-1}\langle\boldsymbol{e}\rangle\nonumber\\
\langle R_{ij}\rangle &=& \frac{\langle e_i^{+}\rangle - \langle e_i^{-}\rangle}{\Delta\gamma_j}.
\end{eqnarray}

We here reiterate that the averages above are over measurements from
different {\it detection catalogs}, generated by running source
detection and ellipticity measurement on differently sheared images. Shear responses
for individual detected objects, as used in both \cite{SheldonMcal2017} and
\cite{HuffMcal2017}, were not calculated.  Doing so would require matching the
lists of detections found on the different sheared images, so that finite
differences for each object could be formed.  This act of matching would
introduce the very shear-dependent object detection biases we wish to
calibrate.  We will discuss the implications of this fact for the analysis of
imaging surveys in \S \ref{sec:wavg}.

\subsection{\textsc{Metadetection} Results for Simulated Galaxy Pairs}
\label{sec:mdetpairs}

In Figure~\ref{fig:pairbias} we show results for the simulations of galaxy
pairs, now including detection in the \mcal\ process. The blue filled circles
represent the case where deblending is performed using MOF. The green plus
signs represent the case where no deblending was performed. For the case
without deblending,  we further simplified the process: we calculated
simple weighted moments at the position determined by \sx\ using a fixed weight function
with full-width at half maximum 1.2 arcsec, without any correction for the PSF.

In both cases the bias is greatly reduced, with significant bias seen only at
the special separation of 1.5 arcsec, where the two objects are detected as one
object by \sx\ in half of the cases. This demonstrates that the bias we see is
not primarily due to the process of deblending itself, but rather
shear-dependent detection effects. The remaining biases at 1.5 arcsec tend to
be different sign for the deblended and non-deblended cases, which shows there
is a qualitative difference in how the two measurements respond to the shear.
As we will show below, we find no significant net bias for more realistic DES
and LSST-like images where the typical separation of galaxies is not at a
special value of maximum detection ambiguity.

\subsection{\textsc{Metadetection} Results for Simulations with Representative Galaxy Density and Noise}
\label{sec:res:constpsf}

We show results for DES-like and LSST-like surveys in Table~\ref{tab:shearmeas}.
We have used a constant PSF and constant shear for these simulations.
Note we here only included \mdet\ with Gaussian weighted moment ellipticities,
primarily to reduce the total CPU time used for the tests.

We find that in all cases our \mdet\ shear measurements are unbiased up to
second-order shear effects \citep[we expect a bias of a few parts in 10000 for
shears of 0.02, see][]{SheldonMcal2017}. This conclusion holds despite the
extensive blending of the object images and the large source detection effects
we documented above. They also meet or exceed the requirements for analyzing an
LSST-like survey \citep[e.g.,][]{huterer2006}. Finally, note that we have also
shown results for an LSST-like survey where the number density of objects is
approximately twice that expected from the actual survey.  Even at these higher
densities we find no increase in the shear bias.

\subsection{Testing the Physical Assumptions Behind \mdet}
\label{sec:mdetphys}

Here we address a key physical assumption made by \mdet, namely that the space
between all objects is sheared coherently.   With \mdet\ we shear the entire
image, so the space between objects is sheared as well as their shapes, and this
is completely coherent across the image.  In \S \ref{sec:res:constpsf} we
showed that, when the shear in the simulation matches this procedure exactly,
we calculate the response accurately.

Real data, however, typically contains images of objects sheared by different
amounts at different redshifts, and can be thought of as a sum of a series of
constant, but differently sheared images.  In such an image, the shearing is
not completely coherent.  Variable shear itself is not a source of bias for
\mdet; the formalism presented in \citep{SheldonMcal2017} recovers the mean
shear or other ensemble statistic for a population.  But because the shearing
of the space between objects in real data is not completely coherent, we may
expect that the part of the \mdet\ response associated with {\em detection} is
slightly biased.

\begin{figure*}
    \begin{center}
        \includegraphics[width=\textwidth]{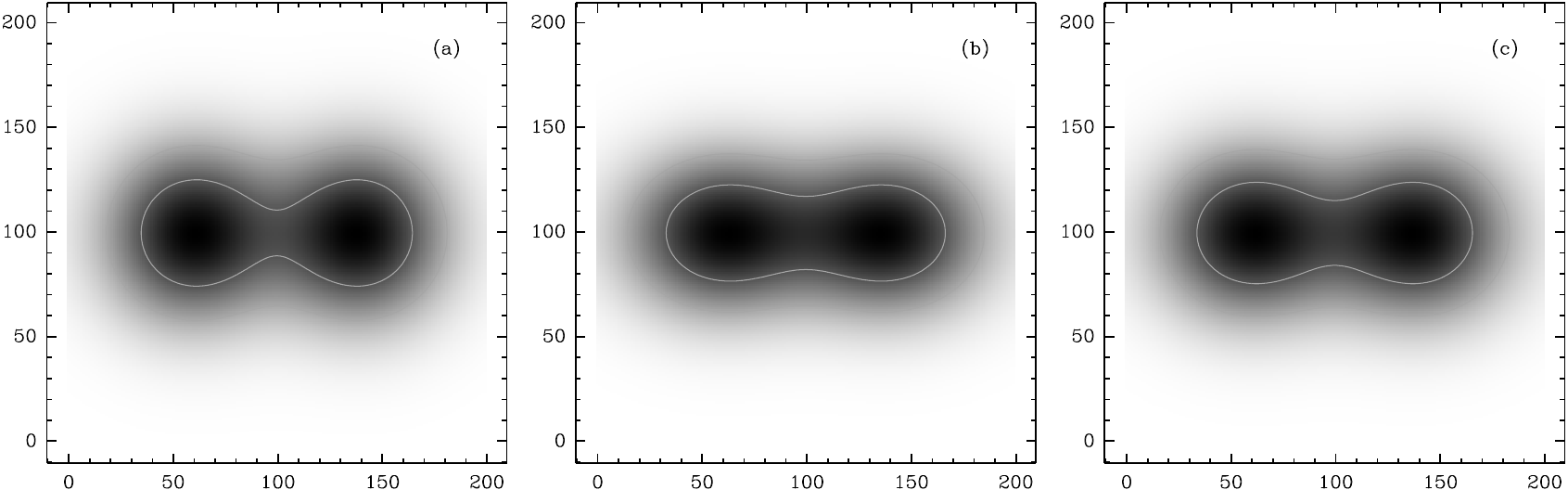}

        \caption{Same as Figure~\ref{fig:toy} but the shear was applied to the
        objects without shearing the space between them. This models the extreme
        and unphysical case where two objects are in line-of-sight projection but
        sheared completely independently.  The contours changed less after shearing
        in this case as compared to the contours in Figure~\ref{fig:toy}.  This
        case differs from the \mdet\ process, in which the entire image is sheared,
        including the space between images.  \label{fig:toynoscene} }
    \end{center}

\end{figure*}

In Figure~\ref{fig:toynoscene} we show a toy example, similar to
Figure~\ref{fig:toy}, demonstrating the extreme and unphysical case of two
objects that are in line-of-sight projection but sheared completely
independently.  We did not allow the space between objects to be sheared. The
contours of constant surface brightness differ less after shear than those
shown in \ref{fig:toy}.  We find this intuitive, because the relative separation
between objects does not change.  The \mdet\ process of shearing the full
image, which does move the positions of objects, will over-predict the response
in this case.

In general, we expect a larger bias due to this effect for surveys with more
object blending, which scales with object density and PSF size.  In order to
obtain an upper bound on this effect for real surveys, we made a simple
modification to our simulations with realistic galaxy density and noise.  When
building them, rather than shearing the full scene to impart the true shear to
the image, we sheared each object individually and then added it to the image,
without any change in the object position. This modification leaves the space
between objects unsheared, which is maximally different from what happens
during the \mcal\ image shearing process.

Results are shown in the bottom rows of
Table~\ref{tab:shearmeas}. We found small residual biases in this case, of
order $\sim-0.3\%$. The fact that we find more bias for a DES-like survey than
an LSST-like survey might be explained by the large PSFs in the DES-like survey
and the smaller pixel scale in the LSST-like survey. However, some of the
DES-like results are puzzling. In particular, the trend with object density in
the DES-like surveys appears to be opposite our naive expectation. We do not
fully understand this effect. While predicting the expected level of this bias
for experiments such as such as DES and LSST is essential, it is beyond the
scope of this work.

\section{Handling PSF Variation}
\label{sec:psfvar}

In order to apply \mdet\ to a real, multi-band, multi-epoch survey like the DES
or LSST, we must address realistic levels of PSF variation, missing data, and
non-trivial WCS transformations. We will address missing data and WCS issues in
a future work, currently in preparation. Here we address the issue of PSF
variation, which is technically the most challenging because \mdet\ requires
deconvolution by the PSF over relatively large regions of sky.

The deconvolution of a spatially varying PSF formally requires a spatially
varying kernel, the implementation of which would be computationally
challenging.  We instead adopted efficient FFTs for deconvolution, which require
a constant kernel.  For this kernel, we chose to use the PSF associated with
the center of the final coadd image, which is thus systematically wrong at
other locations and will necessarily produce a bias in the recovered shear.
However, we will show that the image coadding process used in a realistic
multi-epoch survey results in a more uniform PSF in the final coadd, and
sufficiently reduces associated biases.

For these simulations, we used a simple population of galaxies that have
exponential profiles with a half light radius of 0.5 arcsec. All of the objects
were round and were rendered at a signal-to-noise ratio greater than 20. These
simulations had very low ellipticity noise and so could reach a high precision
with a relatively small number of images. We created the images for the
simulation by coadding a number of images with random, variable PSFs.  \mdet\
was then performed using the coadd of PSF models from the input images, with
the PSF from each input image generated at the location of the center of the
final coadd.

In order to place an upper bound on this effect, we created a variable PSF
model that had significantly more variation than we expect in real data. See
Appendix~\ref{app:pspsf} for details.  Using a single PSF realization from
Appendix~\ref{app:pspsf}, without coadding, we found a multiplicative bias of
$-0.0065 +/- 0.00044$.  However, when coadding thirty of these models, which is
the expected number of epochs for three bands in the final DES data set, we
found a multiplicative bias of only $-0.00035 +/- 0.00037$.  For LSST many more
epochs will be available for coadding.   While this test is not conclusive, we
expect that in a realistic survey scenario, PSF variation will not be a
fundamental limitation for \mdet.

\section{Implications for the Analysis of Imaging Surveys} \label{sec:wavg}

The fact that five separate catalogs must be used without any attempt at
matching them (see \S \ref{sec:mitigate}), has implications for using \mdet\ in
the analysis of surveys.  Typically a single reference or ``gold'' sample of
objects is constructed and used for all subsequent calculations, such as
calculating redshift distributions.  With \mdet, multiple such catalogs, one
for each artificial shear, must be constructed in order to understand the
response of summary statistics to shear.  Specifically, the process of
selecting objects, such as removing objects with low signal-to-noise ratio or
sorting objects into redshift bins, must be repeated using the measurements made on
sheared images in order to include selection effects \citep{SheldonMcal2017}.

When calculating sums and averages for other, non-shear quantities, such as a
redshift distribution, the appropriate object-by-object weight is the shear
response \citep{SheldonMcal2017}.  But individual responses cannot be
calculated with \mdet, because this would require matching the catalogs
produced from the different artificially sheared images, which would introduce
shear-dependent selection effects.  It may be possible to derive an appropriate
mean weight using ensemble statistics.  For example, one could define fine
redshift bins (different from the bins used for tomography) and calculate the
mean response in those bins using the separate sheared versions of the
catalogs. This could then be interpolated to provide weights for objects when
constructing the redshift distribution in tomographic bins.  We will study this
issue in more detail in a future work.

\section{Summary}\label{sec:conc}

In this work, we explored how \mcal\ weak lensing measurements perform in
scenarios where the images of objects overlap and the detection of objects can
be ambiguous. These conditions will characterize all future weak lensing
surveys, especially those executed beneath the atmosphere where PSF smearing
significantly increases the blending of images, so accurate performance in this
regime is critical. We showed that \mcal\ has many percent biases that get worse
as the degree of blending increases, even when objects are deblended.  We then
demonstrated that we can accurately eliminate these biases by including
the detection of objects in the \mcal\ process, even for an LSST-like survey.
We call this technique \mdet.

We tested an important assumption of \mdet, that the space between objects in
an image is sheared coherently, an assumption that does not perfectly match
real data.  We placed an upper bound on the bias associated with this effect at
a few tenths of percents for future surveys. More detailed study will be needed
to predict the actual effect for specific data sets.

In future work we must address a number of technical challenges associated with
implementing \mdet\ on real data.  We must run \mdet\ over relatively large
images ($\sim1-2$ arcminutes on a side) in order efficiently capture
ambiguities in the detection of objects.  Detection is less efficient in
smaller images, because the larger perimeter to area ratio results in a higher
fraction of objects, including blends, near the edge.  Also, over these large
scales, the assumption that the PSF and world coordinate system (WCS)
transformations are approximately constant is incorrect, complicating the
application of an artificial shear. We have shown in this work that PSF
variation is unlikely to be a problem when coadding many tens of images.
However, in order to reconstruct accurate PSF models for the coadd, the input
images can have no edges within the coadd region. This requirement means some
images must be left out of the coadd process, resulting in some loss of some
precision \citep{ArmstrongCoadd}. Finally, we should be able to handle
non-constant WCS transformations by coadding the images into a nearly constant
WCS, but this procedure remains to be tested.

Additional issues arise from masking. Large regions of images can have
non-trivial masking patterns due to stellar diffraction spikes, streaks from
moving objects, cosmic rays, etc. In the implementation presented in this work,
we use fast Fourier transforms (FFTs) to handle convolutions. The FFT does not
permit missing data, so the masked regions must be interpolated in some way.
Care must be taken that this interpolation does not introduce a spurious shear
signal.  An additional compensating mask, rotated at right angles to the real
mask, can be used to restore symmetry to the image \citep{SheldonMcal2017}. The
camera rotations planned for the Rubin telescope will accomplish this
cancellation of systematic effects even more efficiently.  Also, interpolation
correlates the noise in the image, as does the coadding process itself, so the
noise field used for correcting correlated noise effects must also be
propagated through the same coadding and interpolation
\citep{SheldonMcal2017,ArmstrongCoadd}.

Finally, the five separate \mdet\ catalogs must each be incorporated into the
full set of downstream analysis tasks (e.g., photometric redshift estimation,
the construction of summary statistics, etc.) in order to be used for
cosmological constraints (see \S \ref{sec:wavg} for details). This should be
straightforward, but it does require a small change to how shear data are
analyzed.

% The procedures for doing this cannot match the
% catalogs to each other or any external catalog. Any matching of this nature would
% reintroduce detection biases. The procedures also must apply the same
% selection criterion to each of the five catalogs in order to properly measure
% the shear response. These restrictions may have important downstream effects on
% the analysis.

The degree to which these technical challenges can be overcome will ultimately
determine the accuracy of \mdet\ when used to analyze imaging survey data.

\section*{Acknowledgments}

ES is supported by DOE grant DE-AC02-98CH10886, and MB is supported by DOE
grant DE-AC02-06CH11357.  We gratefully acknowledge the computing resources
provided on Bebop, a high-performance computing cluster operated by the
Laboratory Computing Resource Center at Argonne National Laboratory, and the
RHIC Atlas Computing Facility, operated by Brookhaven National Laboratory.
This work also used resources made available on the Phoenix cluster, a joint
data-intensive computing project between the High Energy Physics Division and
the Computing, Environment, and Life Sciences (CELS) Directorate at Argonne
National Laboratory.

%\bibliographystyle{mnras}
%\bibliography{references}
%\bibliography{apj-jour,references}
%\bibliographystyle{apj}
\bibliographystyle{aasjournal}
\bibliography{references}

\appendix

\section{Fast Approximate Variable PSF Models}\label{app:pspsf}

In this work we used a fast, approximate variable PSF model. This model eases the
computational requirements for the simulations while also retaining the
essential features of realistic PSF variation. In this appendix, we present
the model and verify its statistical properties against more realistic PSF models.

We began with the results of \citet{heymans2012}. They fit the \vonkarman model
of atmospheric turbulence
\begin{displaymath}
  P(\ell) \propto \left(\ell^{2} + \frac{1}{\theta_{0}^2}\right)^{-11/6}
\end{displaymath}
to images with high stellar density. Here $\theta_{0}$ is the outer scale of
turbulence. \citep{heymans2012} find that $\theta_{0}\approx3$ arcmin.
We further added an additional Gaussian truncation of the power
\begin{displaymath}
  P_{trunc}(\ell) \propto P(\ell)\exp\left(-\ell^2r^{2}\right)
\end{displaymath}
at a scale of $r=1$ arcsec in order to reduce the level of resulting
PSF variation. Below we show that even with this modification, our models
still have more power than a realistic model for a survey, making them useful
for providing upper limits on the effects of PSF variation.

Using this model, we seeded equal amounts of E- and B-mode power on a grid of
$128\times128$ cells using random phases. Each cell of the grid was one
arcsec in size. We normalized the overall ellipticity variance to $0.10^2$. We then used
the $g_1$ and $g_2$ components of this model to set the variation of the ellipticity of
the PSF. We drew the mean ellipticity for each image from a Gaussian distribution of
variance $0.10^2$. Note that we also bound the total ellipticity to at most 0.5.
We modeled the PSF profile as a Moffat with shape parameter $\beta=2.5$.
The size of the Moffat profile was set to be proportional to $\mu^{-3/4}$,
where $\mu$ is the magnification computed from the power spectra realization. The
proportionality constant was drawn randomly from a log-normal model with
scatter 0.1 arcmin and a central value set so the final PSF size mimicked a DES-like
survey, with focal plane averaged FWHM $\sim1.1$ arcsec.

\begin{figure*}
    \begin{center}
        \includegraphics[width=0.7\textwidth]{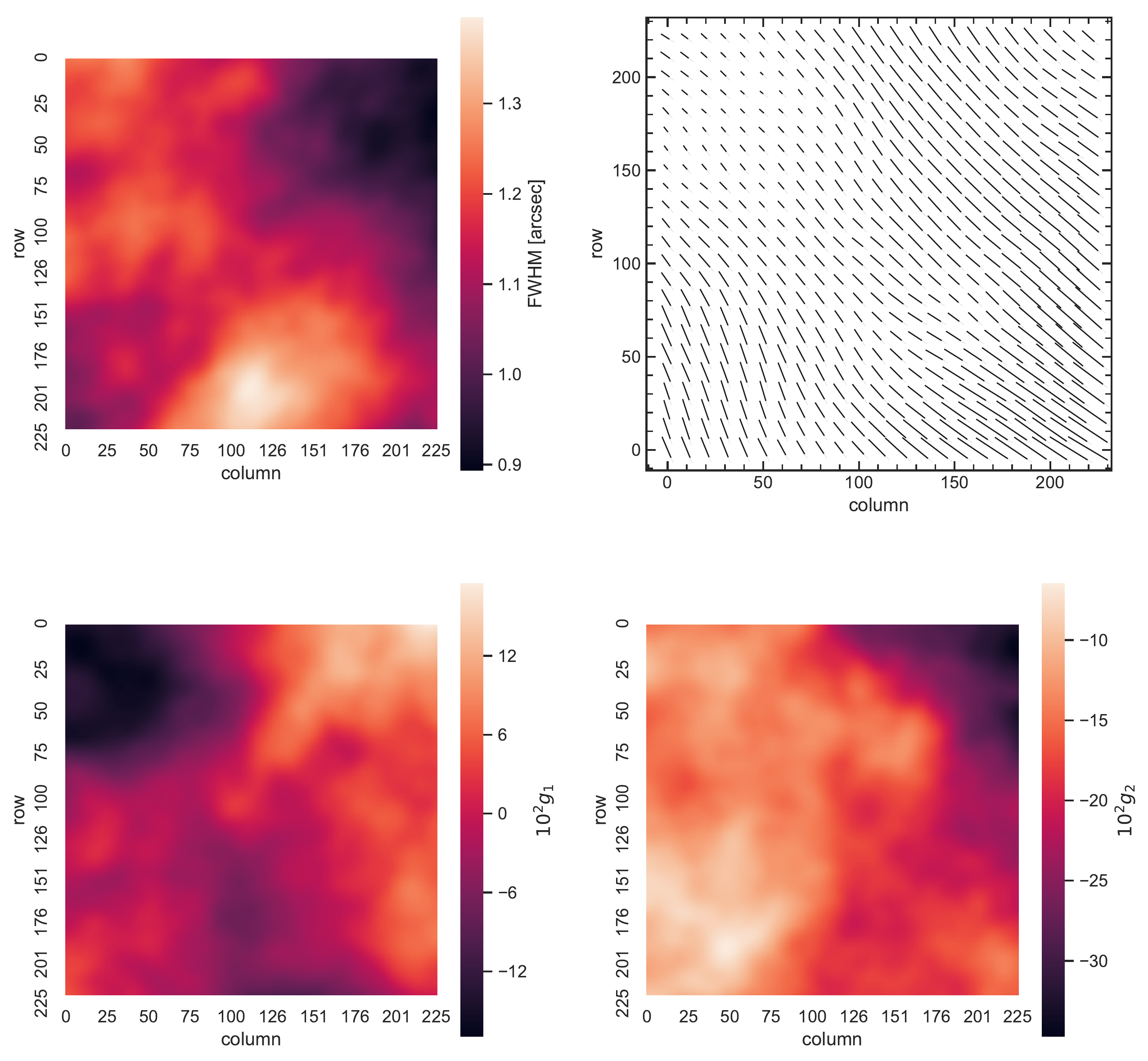}
        \caption{
            Variable PSF model statistics for a DECam-like exposure. The top-left
        panel shows the variation in the FWHM in arcseconds. The top-right panel
        shows a visualization of the PSF ellipticity variation. The bottom-left panel shows
        the variation in the $1$-component of the PSF ellipticity. The bottom-right panel
        shows the variation in the $2$-component of the PSF ellipticity. The variation in
        this model is $\gtrsim10\times$ larger than the typical PSF variation for
        either DECam or expected LSST observations. The pixel scale is 0.263 arcsec
        so that each panel is approximately 1 arcmin on a side.
        \label{fig:pspsf}}
    \end{center}
\end{figure*}

\begin{figure}
    \begin{center}
        \includegraphics[width=\columnwidth]{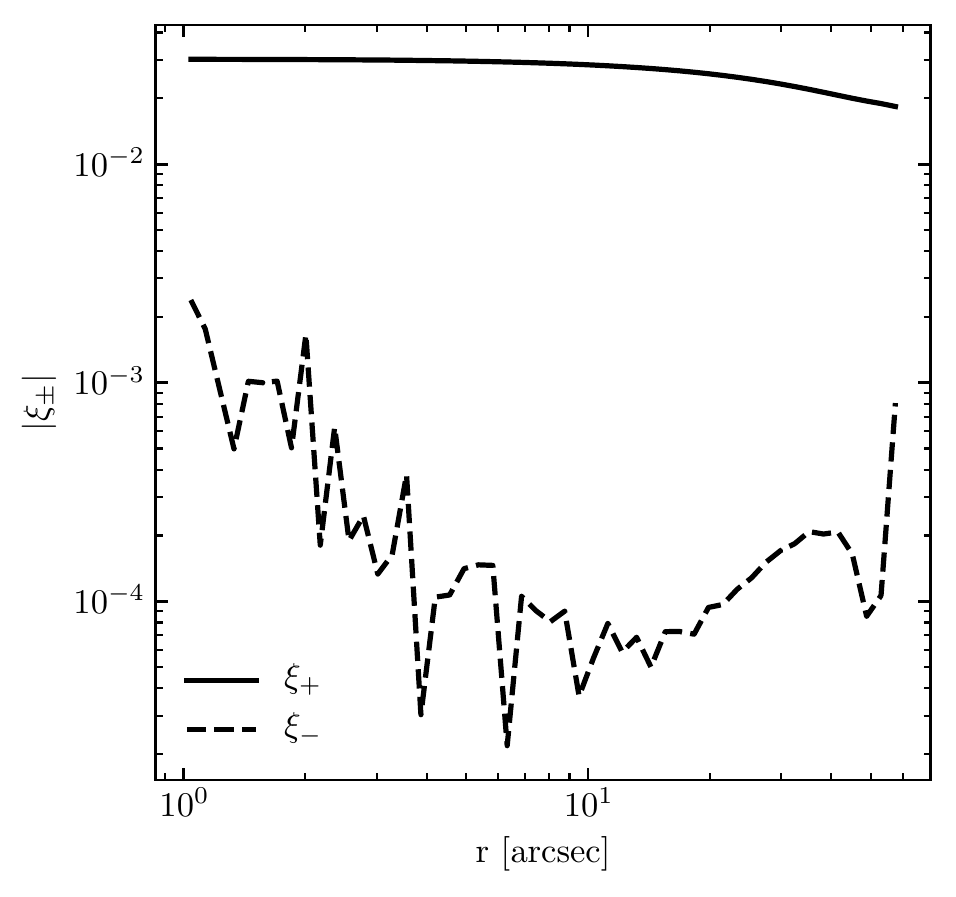}
        \caption{
            Variable PSF model shear correlation functions for a DECam-like exposure. LSST
        is expected to have shear correlation function magnitudes around
        $\sim\!10^{-4}$ \citep{jee2011}.
        \label{fig:psxi}}
    \end{center}
\end{figure}

We show an example PSF for a DES-like survey in Figure~\ref{fig:pspsf}.  Over a
1 square arcminute patch, our approximate models generate PSF ellipticity and size
variation that are $\gtrsim10\times$ that seen in real 90 second exposures
with DECam \citep{DESY1shear}, or the expected variation in a 15 second exposure with
LSST \citep{jee2011} over similar scales. Figure~\ref{fig:psxi} shows the $\xi_{\pm}$ shear correlation
functions averaged over 100 realizations of our models. For comparison, we
expect at most shear correlation function amplitudes of $\sim10^{-4}$ for LSST
\citep{jee2011} and for DESCam 90 second exposures. The DECam models were generated
using the methods of \citet{jee2011} but for DECam-like environmental conditions. For
the optical contributions to the PSF, we use a set of randomly drawn
aberrations (similar to GREAT3 \citep{great3}), but with values more typical of
DECam observations\footnote{\url{https://github.com/GalSim-developers/GalSim/blob/releases/2.1/examples/great3/cgc.yaml}}.

%\bsp
%\label{lastpage}
\end{document}